\documentstyle[preprint,eqsecnum,aps,prb]{revtex}

\begin{document}
\title{Lattice coupled first order magnetoresistance transition in an A-type
antiferromagnet: Pr$_{0.46}$Sr$_{0.54}$MnO$_3$ }
\author{R. Mahendiran$^{1}$, C. Martin$^{1}$, A. Maignan$^{1}$,M. Hervieu$^{1}$, B.
Raveau$^{1}$ and L. Morellon$^{2}$, C. Marquina$^{2}$, B. Garc\'{i}a-Landa$%
^{2}$, and M. R. Ibarra$^{2}$}
\address{$^1$Laboratoire CRISMAT, ISMRA, Universit\'{e} de Caen, 6 Boulevard du
Mar\'{e}chal Juin, 14050 Caen-Cedex, France.\\
$^2$ Departamento de Fisica de la Materia Condensada-ICMA, Universidad de
Zaragoza-CSIC, 50009 Zaragoza, Spain.}
\date{\today}
\maketitle

\begin{abstract}
We investigated magnetic, magnetotransport and magnetostriction properties
of the A-type antiferromagnet Pr$_{0.46}$Sr$_{0.54}$MnO$_{3}$ which
undergoes a first order paramagnetic-antiferromagnetic transition below T$%
_{N}$ = 210 K while cooling and T$_{N}$ = 215 K while warming. The zero
field ($\mu _{0}$H = 0 T) resistivity shows a sudden jump at T$_{N}$ \ and a
small bump around T$_{max}$ = 220 K (%
\mbox{$>$}%
T$_{N}$). T$_{N}$ shifts down and T$_{max}$ shifts up with increasing $\mu
_{0}$H. Magnetoresistance as high as -45-57 \% at 7 T is found in the
temperature range 180 K-230 K. Isothermal measurements indicate that \ the
field induced antiferromagnetic to ferromagnetic transition below T$_{N}$ is
accompanied by a rapid decrease of the resistivity but increase of volume ($%
\Delta $V/V = +0.25 \% at 180 K and 13.7 T). \ This lattice coupled
magnetoresistance transition is suggested due to the field induced
structural transition from the low volume orthorhombic to the high volume
tetragonal structure.
\end{abstract}


\bigskip

Extensive studies following the discovery of a magnetic field induced
destruction of the antiferromagnetic insulating phase in Pr$_{0.5}$Sr$_{0.5}$%
MnO$_{3}$ revealed that the low temperature insulating phase is actually a
two dimensional ferromagnetic metal but with successive ferromagnetic planes
coupled antiferromagnetically \cite{Kuwahara,Damay}. This peculiar type of
antiferromagnetism known as A-type antiferromagnetism is caused by two
dimensional ordering of the Jahn-Teller split e$_{g}$ -d$_{x^{2}-y^{2}}$
orbitals of Mn$^{3+}$ ions. When heated from the low temperature, the
interplanar antiferromagnetic coupling becomes weak and Pr$_{0.5}$Sr$_{0.5}$%
MnO$_{3}$ transforms into a ferromagnetic metal in the temperature range
145- 270 K before changing into a paramagnetic insulator. There are
indications that ferromagnetic domains from nano to micron size coexist with
the antiferromagnetic phase even far below the Neel temperature in Pr$_{0.5}$%
Sr$_{0.5}$MnO$_{3}$ by NMR\cite{Allodi} and magnetic studies\cite{Mahi}.
With increasing Sr content the strength of A-type antiferromagnetic
interaction increases at the expense of the ferromagnetic interaction and
finally A-type antiferromagnetism gives away to the C-type
antiferromagnetism in Pr$_{0.3}$Sr$_{0.7}$MnO$_{3}$\cite{Martin1,Hervieu}.
The A-type antiferromagnetism is so far not observed for x 
\mbox{$<$}%
0.5 to in any of the RE$_{1-x}$AE$_{x}$MnO$_{3}$ (RE is a trivalent rare
earth ion and AE is a divalent alkaline earth ion) series. Although there
are few electrical transport studies in the A-type antiferromagnetic
compounds away x = 0.5\cite{Martin1,Okuda,Moritomo}, there is hardly any
study on magnetotransport in the A-type antiferromagnetic compounds away
from half doping\cite{Martin1,Okuda} . Kuwahara et al\cite{Okuda} found \
that the out plane resistivity in Nd$_{0.45}$Sr$_{0.55}$MnO$_{3}$ is 4
orders of magnitude higher than the in plane value at 45 mK and the in plane
resistivity at $\mu _{0}$H = 10 T and 45 mK is an order of magnitude lower
than the out of plane resistivity. However, it is not known whether lattice
distortion plays any role in the magnetoresistance of the A-type
antiferromagnet. In order to investigate this aspect we have chosen Pr$%
_{0.46}$Sr$_{0.54}$MnO$_{3}$ which is an A-type antiferromagnet and shows a
large magnetoresistance close to the Neel temperature\cite{Martin}. To
clarify this aspect we carried out magnetostriction measurement on Pr$_{1-x}$%
Sr$_{x}$MnO$_{3}$ (x = 0.54, 0.55). \ We also investigated Pr$_{0.45}$Sr$%
_{0.55}$MnO$_{3}$ \ but a detailed study was discarded because a magnetic
field \ more for $\mu _{0}$H = 14 T which is beyond our experimental limit
was needed to induce an antiferromagnetic to ferromagnetic transition.

\bigskip

We measured four probe resistivity of the polycrystalline Pr$_{0.46}$Sr$%
_{0.54}$MnO$_{3}$ over a wide temperature (10 K- 320 K) and field ( 0 T $%
\leq $ H $\leq $ 7 T) range using a Quantum Design Physical Properties
Measuring System. Magnetic measurements were carried out using a Quantum
Design SQUID magnetometer. In the ac susceptiblity measurement the amplitude
of ac signal was 3 Oe and the frequency was 100 Hz. Magnetostriction
measuremnts using the strain gauge method in a pulsed field up to13.7 T was
measured at University of Zaragoza, Spain. We measured both parallel ($%
\lambda _{par}$) and perpendicular ($\lambda _{per}$) magnetostrictions with
field parallel and perpendicular to the measuring directions respectively.
The volume magnetostriction of randomly oriented polycrystallites is
determined through the relation $\omega $ = $\lambda _{par}$+2$\lambda
_{per} $

\bigskip

Fig. 1 (a) shows the real ($\chi $') and imaginary ($\chi $'') parts of the
ac susceptibility while warming from 10 K. Both $\chi $' and $\chi $''
exhibit a peak around T$_{N}$ = 215 K due to A-type antiferromagnetic to
paramagnetic transition. The paramagnetic to antiferromagnetic transition
while cooling occurs at T$_{N}$ = 210 K but the data is not shown for
clarity. The antiferro-paramagnetic transition is accompanied by
orthorhombic (Fmmm) to tetragonal (I4/mcm) structural transition as revealed
by neutron diffraction study\cite{Martin2}. The inverse susceptibility is
shown on the right scale. The low temperature anomaly in the inverse
susceptibility (1/$\chi $') around 50 K is possibly related to ordering of \
the Pr moments and above T$_{N}$, 1/$\chi $'obeys the Curie-Weiss law with a
positive Weiss temperature $\theta _{p}$ = 230 K and an effective moment of P%
$_{eff}$ = 7.1837 $\mu _{B}$. The experimentally observed value of P$_{eff}$
is higher than the theoretically calculated value of P$_{eff}$ =$\sqrt{0.46%
\text{P}_{\Pr^{3+}}^{\text{2}}+0.46\text{P}_{\text{Mn}^{3+}}^{\text{2}}+0.54%
\text{P}_{\text{Mn}^{4+}}^{\text{2}}}$ = 5.004 $\mu _{B}$ where P$%
_{\Pr^{3+}}^{2}=$ 12.816, P$_{\text{Mn}^{3+}\text{\ }}^{2}$= 24, P$_{\text{Mn%
}^{4+}}^{2}$ = 15. Judging from the earlier magnetization and resistivity
data in Pr$_{1-x}$Sr$_{x}$MnO$_{3}$ series\cite{Martin} a short range
ferromagnetic ordering with T$_{C}$ $\approx $220 K is not an unlikely
possibility in x = 0.54. The closeness of T$_{C}$ to T$_{N}$ makes it
difficult to distinguish in x = 0.54 from the susceptibility data alone. A
strong ferromagnetic correlations exist much above the T$_{C}$ as suggested
by the high effective magnetic moment value. When the strength of the
external magnetic field increases, ferromagnetic correlations becomes much
stronger and T$_{C}$ shifts up in temperature. This is reflected in Fig.
1(b) which shows the inverse dc susceptibility (H/M) while field cooling for
different H values over 320K-160 K. The curves for different H do not match
in the paramagnetic region over a wide temperature range (200 K- 300 K). The
short range ferromagnetic ordering is also reflected in the M(H) isotherms
in Fig. 1(c) at T = 215 K, 220 K, and 225 K. As T is reduced \ below 210 K,
antiferromagnetic correlations overcome ferromagnetic contributions. At 210
K, M(H) increases initially increases linearly with the field but then
raises rapidly around $\mu _{0}$H$_{c}$ = 1.2 T. The rapid increase of M(H)
signals antiferromagnetic to ferromagnetic transition. The transition is
first order as suggested by the hysteresis behavior. As T decreases further $%
\mu _{0}$H$_{c}$ increases to rapidly to 4.0 T at 200 K.

\bigskip

Fig. 2(a) shows temperature dependence of the resistivity $\rho $(T) for H.=
0 T and 7 T. $\rho $(T) increases gradually as T is lowered below 320 K and
exhibits a bump around T$_{max}$ = 220 K as shown in the inset . Then, $\rho 
$(T) jumps up around T$_{N}$ = 210 K and then continues to increase.. The
value of $\rho $(T) at 4.2 K is very small ($\approx $30.0 m$\Omega $ cm)
which is slightly larger than the in plane resistivity at 4.2 K ($\approx $
10 m$\Omega $ cm) of the single crystal A-type antiferromagnet Nd$_{0.45}$Sr$%
_{0.55}$MnO$_{3}$\cite{Okuda}. The weak bump at T$_{max}$ is caused by the
short range ferromagnetic order\cite{Martin}. The T$_{max}$ is shifted up to
more than 300 K and T$_{N}$ is shifted down to 184 K under $\mu _{0}$H = 7
T. Fig. 2(b) shows $\rho $(T) curves while cooling and warming under a field
for different values of H over a limited temperature range. The $\rho $(0T)
curve exhibits hysteresis of 5 K width during the paramagnetic (tetragonal)
to antiferromagnetic (orthorhombic) while cooling and the reverse transition
upon warming. We take the middle point of the hysteresis region as the Neel
temperature for convenience. When the field is increased T$_{N}$
systematically shifts down and T$_{max}$ shifts up as shown in Fig. 3. The
widening of difference between T$_{C}$ and T$_{N}$ is reflected in the
metallic like resistivity behavior in between T$_{C}$ and T$_{N}$. The
applied magnetic field strongly reduces the resistivity in between T$_{C}$.
and T$_{N}$.

\bigskip

Fig. 4 (a) shows the field dependence of the magnetoreistance MR = [$\rho $%
(H)-$\rho $(0)]/$\rho $(0) at selected temperatures. The data were taken
after zero field cooling to a given temperature from T = 300 K. At 215 K,
the magnitude of MR increases rapidly for H 
\mbox{$<$}%
1 T and then gradually for higher fields. The data taken at 220 K, 225 K and
230 K (not shown here for clarity) are similar to the 215 K. The field
dependence of these data are similar to the MR behavior of the spin
dependent tunnelling between the ferromagnetic grains in polycrystalline
manganites below the ferromagnetic Curie temperature\cite{Hwang}. When T is
decreased below 210 K, MR gradually decreases with H initially and then
rapidly around $\mu _{0}$H$_{C}^{+}$ before levelling off at high fields. At
205 K, $\mu _{0}$H$_{C}^{+}$ = 2.2 T during field increasing part which
closely correlates with $\mu _{0}$H$_{C}^{+}$ found in the M(H) data in Fig.
1(c). In the A-type antiferromagnetic state, metallic conduction is two
dimensional in absence of magnetic field. When the interplanar
antiferromagnetic coupling is destroyed by a strong magnetic field, the
charge transport changes into three dimensional in character which results
in a rapid increase in the value of the MR. The field induced
antiferromagnetic to ferromagnetic transition is first order with a constant
hysteresis width of 1.5 T. The maximum magnetoresistance at 7 T is -50-55 \%
which is comparable to other manganites.

\bigskip

Is the MR due to changes in the spin sector alone ? To understand this
aspect, we measured magnetostriction isotherms which are shown in Fig. 4(b).
We find a giant volume expansion with hysteresis in the temperature range
205 K-170 K. This behavior is opposite the volume contraction under a
magnetic field found in most of the manganites investigated earlier\cite
{Ibarra}. The volume change is rather small when the sample is in the AF
state but shows a sudden jump during the field induced antiferromagnetic to
ferromagnetic transition. The critical field $\mu _{0}$H$_{C}^{+}$= 2.25 T
at 205 K is in close agreement with the MR\ and M(H) data. After the
completion of the AF-FM transition $\Delta $V/V shows a gradual variation
with further increase in H.is gradual. At 210 K (i.e., T$\approx $ T$_{N}$) $%
\Delta $V/V is small and it attains the maximum value of 0.25 \% at 180 K
and then decreases to 0.16 \% at 170 K. This is because the maximum
available field of 13.7 T is insufficient to complete the AF-FM transition.
What is the origin of the observed volume expansion under the magnetic field
?. The A-type AF order is coupled to the d$_{x^{2}-y^{2}}$ orbital ordering.
Orbital ordering creates quadrupole moments which can interact with the
lattice through electron phonon coupling \cite{Khomskii}. The orbital
ordering is also destroyed together with the AF ordering for H $>>$ $\mu
_{0} $H$_{C}$. This triggers a field induced structural transition from the
orthorhombic to possibly a tetragonal \ structure. This possibility is more
likely given the fact the antiferromagnetic phase has lower volume than the
paramagnetic phase as revealed by the volume thermal expansion data shown in
Fig. 5. The antiferromagnetic phase has orthorhombic structure and the
paramagnetic phase has tetragonal structure\cite{Martin2}. The volume
thermal expansion decreases by $\approx $ 0.18 \% in between 220 K and 200
K.as the temperature is decreased and the structural transition shows
hysteresis. If there were no magnetic and structural transition, $\Delta $%
V/V would have followed the Gr\H{u}neissan behavior shown as a fit to the
high temperature data. The volume expansion under the magnetic field has
contributions one from volume change due to magnetic field induced
structural transition and another from volume change due to spontaneous
magnetostriction of \ the field induced ferromagnetic phase. At 190 K the
experimentally observed volume is 0.21 \% lower than the Gr\H{u}neissan
value whereas the volume magnetostriction is 0.25 \%. This small difference
is due the contribution from spontaneous magnetostriction of the
ferromagnetic phase.. Thus, our study indicates that the magnetotransport in
the A-type AF Pr$_{0.46}$Sr$_{0.54}$MnO$_{3}$ is strongly coupled to the
structural transition under the field. It will be interesting to know
whether\ a large first order magnetoresistance behavior can be ever observed
in \ a manganite without a change in lattice distortion or structural
transition.

\bigskip

In conclusion, we have shown the A-type AF Pr$_{0.46}$Sr$_{0.54}$MnO$_{3}$
exhibits a first order magnetoresistance behavior below the Neel temperature
which involves not only spin and charge but also the lattice. We found that
the field induced antiferromagnetic to ferromagnetic transition is
accompanied by a giant volume expansion which is suggested to the field
induced structural transition from the low volume orthorhombic to the high
volume tetragonal structure. First order spin-charge-lattice coupled
transition is also of great interest to study giant magnetocaloric effect 
\cite{Pecharsky} and future research on our compound need to be focussed on
this aspect.

\bigskip

{\bf Acknowledgments}

\bigskip

R. M thanks MENRT (France) and Ministrio de Ciencia y Cultura (Spain) for
financial assistance.

\newpage

\begin{center}
{\bf Figure captions}
\end{center}

\begin{description}
\item[Fig. 1]  :Left scale; Temperature dependence of the real ($\chi $')
and imaginary ($\chi $'') parts of \ the ac susceptibility of Pr$_{0.46}$Sr$%
_{0.54}$MnO$_{3}$. Right scale 1/$\chi $' versus temperature. The
Curie-Weiss fit is shown by the thick line. (b).The inverse dc magnetic
susceptibilty for different values of H in a limited temperature range while
field cooling from 320 K (c). Isothermal magnetization data..

\item[Fig. 2]  :Temperature dependence of \ the resistivity \ $\rho $(T) at $%
\mu _{0}$H = 0 T and 7 T over 320 K-10 K range while field cooling. Inset
shows the bump around 220 K ( 
\mbox{$>$}%
T$_{N}$) in a enlarged scale. (b) $\rho $(T) for different $\mu _{0}$H
values in a limited temperature range (160 K- 320 K).

\item[Fig. 3]  : Field dependence of the Neel temperature (T$_{N}$) and the
temperaure T$_{max}$ corresponding to the bump in the resistivity above T$%
_{N}$.

\item[Fig. 4]  : Field dependence of \ the isothermal (a) magnetoresistance
and (b) volume magnetostriction.

\item[Fig. 5]  : Volume thermal expansion of Pr$_{0.46}$Sr$_{0.54}$MnO$_{3}$%
. The structural determination is from the neutron diffraction study (Ref.
9) The hysteresis region is marked by the vertical dashed lines.
\end{description}

\end{document}